\documentclass[aps,prx,twocolumn,showpacs,showkeys,noeprint,longbibliography]{revtex4-1}
\usepackage[utf8x]{inputenc}
\usepackage[T1]{fontenc}
\usepackage{cprotect}
\usepackage{fvextra}
\usepackage{textcomp}
\usepackage{amsmath}
\usepackage{latexsym}
\usepackage{float}
\usepackage{accents}
\usepackage{amssymb}
\usepackage{graphicx}
\usepackage{hyperref}
\usepackage{subfigure}
\usepackage{bbold}
\usepackage{xcolor}
\usepackage{url}
\usepackage{booktabs,tabularx,dcolumn}
\def\ba{\begin{eqnarray}}
\def\ea{\end{eqnarray}}
\def\be{\begin{equation}}
\def\ee{\end{equation}}
\def\bm{\begin{math}}
\def\me{\end{math}}

\newcommand{\dummy}

\begin{document}
\title{Motion of a Polymer Globule with Vicsek-like Activity: From Super-diffusive to Ballistic Behavior}
\author{ Subhajit Paul$^1$} \email[]{subhajit.paul@itp.uni-leipzig.de}
\author {Suman Majumder$^1$}\email[]{suman.majumder@itp.uni-leipzig.de}
\author{Wolfhard Janke$^1$}\email[]{wolfhard.janke@itp.uni-leipzig.de}
\affiliation{$^1$Institut  f\"{u}r Theoretische Physik, Universit\"{a}t Leipzig, IPF 231101, 04081 Leipzig, Germany}

\date{\today}

\begin{abstract}
~~Via molecular dynamics simulation with Langevin thermostat we study the structure and dynamics of a flexible bead-spring active polymer model  after a quench from good to poor solvent conditions.  The self propulsion is introduced via a Vicsek-like alignment activity rule which works on each individual monomer in addition to the standard attractive and repulsive interactions among the monomeric beads. We observe that the final conformations are in the globular phase for the passive as well as for all the active cases. By calculating the bond length distribution, radial distribution function, etc., we show that the kinetics and also the microscopic details of these \textit{pseudo equilibrium} globular conformations are not the same in all the cases. Moreover, the center-of-mass of the polymer shows a more directed trajectory during its motion and the behavior of the mean-squared-displacement gradually changes from a super-diffusive to ballistic  under the influence of the active force in contrast to the diffusive behavior in the passive case.
\end{abstract}


\maketitle

\section{ Introduction}
\par 
Properties of various biological constituents can be understood under the framework of ``active matter'' models which got significant interest to the statistical physics community in the past few decades \cite{ramasw1,roman,cates1,elgeti1,winkler,shaeb,vicsek,tailleur,chate1,mishra2,loi,vic2,fily,redner,cates2,skdas,das1,paul1,jiang,biswas,holder,kaiser,duman,bianco}. The constituents, so-called ``active particles'', have the ability of their own decision making either by converting their internal energy to work or taking energy from the environment. This leads to self propulsion due to which these objects show directed motion and always remain out of equilibrium. Being ubiquitous in nature, such objects are seen over a very wide range of length scales, from bacteria, sperm, algae, etc. at the microscopic single cell level to flocks of birds, schools of fish, etc. in the macroscopic world \cite{ramasw1,vic2,elgeti1,tailleur,winkler}. Though the governing factors are entirely different, the interesting common feature is that such objects always move in a group and in a coherent way.
\par 
The first minimal model in this direction to describe such collective behavior was by Vicsek \textit{et al.}\ \cite{vicsek}. In this model very simple dynamical rules were used to show the clusters formed by point-like particles. In the last few years another most studied model in literature is a system consisting of active Brownian particles (ABP) \cite{cates1,mishra2,redner,fily,cates2,roman}. In the Vicsek model at every instant a particle changes its direction of motion  by looking at the average direction of its neighbors. On the other hand, a system with ABP shows activity induced clustering for completely repulsive interactions among the particles \cite{cates1,cates2,mishra2,fily,roman,redner}. In recent years interest has grown in  modeling active polymers \cite{winkler,holder,duman,kaiser,bianco} which can be visualized as a system of constrained motion of micro-swimmers. They have relevance with various biological objects, e.g.,  bacterial flagellum, microtubules, actin filaments, etc. These filamentous objects can deform or bend and play major roles in determining the motion and shape of cells to which they belong \cite{hakim}. As a specific example, the microtubules that are part of the cytoskeletons in eukaryotic cells are like linear polymers made up of tubulin proteins. They help in maintaining the shape of a cell and its membrane and also work as cargo by taking part in cell motility, intracellular transport, etc. supported by some kind of binding or attachment proteins, \textit{viz.,} kinesin, dyenin, etc. \cite{howard} Thus understanding the dynamics as well as conformational properties of active filaments can help us in elucidating  some biological mechanisms.
\par 
In this regard efforts were mostly directed to understand the properties of active Brownian filaments \cite{winkler,holder,duman,kaiser,bianco}. Such a filament model can be constructed in a straightforward manner by considering the active Brownian particles as monomeric beads and joining them via springs. Focus was mainly to study the collective behavior and pattern formation by such filaments, for which in most of the cases the passive non-bonded monomeric interaction was considered to be a completely repulsive one \cite{holder,duman}.  Recently, via Brownian dynamics simulation of a single active filament in a good solvent, the activity induced conformational changes from coil to globule as well as its enhanced diffusion have been shown \cite{bianco}. In our very recent work \cite{paul2}, upon quenching a flexible polymer from good to a poor solvent condition, we looked at the effect of Vicsek-like alignment activity on its coil-globule transition  with particular focus on the coarsening kinetics. Such coil-globule transition, in the context of a passive polymer, has similarities with the  dynamics of protein folding or chromosome compactification \cite{reddy,shi2018}. For a passive polymer, the monomers can be made ``active'' by some external non-thermal forces. Dynamics of such filaments has been studied in active solvent, with or without hydrodynamic interactions \cite{kaiser1,winkler,chaki,samanta_16}. In experiments active filaments have been designed by joining the chemically synthesized molecules, colloids or Janus particles via DNA strands \cite{biswas,jiang}. Then the activity is introduced via various phoretic effects, i.e., application of light,  electric or magnetic fields. There also it is shown that the activity enhances the diffusive behavior of the polymer chain. 
\par 
Keeping these studies in mind, in this paper, we model an active flexible homopolymer in which the beads follow the Vicsek-like alignment activity rule \cite{vicsek,paul2,paul1,das1}.  The kinetics of the formation of a single globule for the passive limit of the model has been extensively studied in literature with both Monte Carlo and molecular dynamics simulations \cite{halperin,majumder1,majumder3,majumder4,christiansen,byrne,bunin,guo,milchev}. But such studies are much lesser in the context of a single active polymer \cite{bianco,paul2,kaiser}.  In this work we will mainly look at the motion of an active polymer in implicit solvent with particular emphasis on the microscopic structural details of its \textit{pseudo equilibrium} steady state conformations and compare the results with those from its passive limit. 
\par 
The rest of the paper is organized as follows. In Sec. II we discuss the model and methods of our simulations in detail. Section III contains the results followed by the conclusions in Sec. IV.\\

\section{ Model and Methods}

~We consider a model flexible polymer in which the monomer beads are connected via spring-like arrangements. For the active polymer model, self propulsion is added for each bead. Before looking at how the active force is included for the beads, first we discuss the various passive interactions among the beads.
The monomer-monomer bonded interaction has been modeled via the standard finitely extensible non-linear elastic (FENE) potential \cite{milchev, majumder1, majumder3, majumder4} defined as
\begin{equation}
V_{\rm{FENE}}(r) = - \frac{K}{2}R^2 {\rm{ln}} \bigg[ 1- \bigg(\frac{r-r_0}{R}\bigg)^2\bigg],
\end{equation}
where $r_0~(=0.7)$ is the equilibrium bond distance. $K$ is the spring constant which is set to $40$ and $R$ measures the maximum extension of the bonds on both sides of $r_0$, for which the value is chosen to $0.3$.
\par 
~The non-bonded monomer-monomer interaction is modeled via the standard Lennard-Jones (LJ) potential \cite{majumder1, majumder3,das1,paul1}
\begin{equation}\label{lj_pot}
 V_{\rm{LJ}}(r) = 4\epsilon \bigg[\bigg(\frac{\sigma}{r}\bigg)^{12}- \bigg(\frac{\sigma}{r}\bigg)^6\bigg],
\end{equation}
where $r$ is the distance between the monomers and $\epsilon$ is the interaction strength, value of which is set to unity. This measures the energy scale of the system. The length scale of our system is expressed in units of $\sigma$, the diameter of the beads, which is related to $r_0$ as $\sigma= r_0/2^{1/6}$. Following our nonequilibrium study \cite{paul2}, here also we consider both attractive and repulsive forces for the non-bonded interaction among the monomers to ensure a poor solvent condition for the polymer and thus formation of globular conformations.
\par 
While working with the full form of $V_{\rm{LJ}}$, the potential is truncated and shifted at $r_c=2.5\sigma$ for advantages during numerical simulations. In  that case the non-bonded pairwise interaction takes the form
\begin{equation}
  V_{\rm{NB}}(r)=
\begin{cases}
  V_{\rm{LJ}}(r)-V_{\rm{LJ}}(r_c) -(r-r_c)\frac{dV_{\rm{LJ}}}{dr}\Big|_{r=r_c}  r<r_c \,,\\
 0 ~~~~~~~~~~~~~ \text{otherwise}\,,
   \end{cases}
\end{equation}
having similar behavior as $V_{\rm{LJ}}$. 
\par 
 By truncating $V_{\rm{LJ}}$ of Eq.~(\ref{lj_pot}) at its minimum, i.e., at $r_c=r_0=2^{1/6}\sigma$ \big(where $V_{\rm{LJ}}(r_0)=-\epsilon$ and $\big({dV_{\rm{LJ}}}/{dr}\big)_{r=r_0} =0$\big) $V_{\rm{NB}}(r)$ becomes the completely repulsive Weeks-Chandler-Andersen (WCA) potential \cite{wca_71}
\begin{equation}\label{wca_pot}
V_{\rm{WCA}}(r)=
\begin{cases}
V_{\rm{LJ}}(r)+\epsilon~~~~~~~~  r<r_0 \,,\\
0 ~~~~~~~ \text{otherwise}\,.
\end{cases}
\end{equation}
\par
The dynamics of a passive polymer in a poor solvent modeled by $V_{\rm NB}$ is studied via molecular dynamics (MD) simulations \cite{frenkel}. The temperature for the polymer is kept constant by employing the Langevin thermostat \cite{das1,paul1}. 
Thus, for each bead we work with
\begin{equation}\label{langevin}
m_i \ddot{\vec{r}}_i = - \vec{\nabla} U_i - \gamma \dot{\vec{r}}_i + \sqrt{2 \gamma k_B T } \vec{R}_i(t),
\end{equation}
where the mass $m_i~(=m)$ is unity for all the beads,  $\gamma$ is the drag coefficient, which we set $\gamma=1$, and $k_B$ is the Boltzmann constant, value of which 
is also set to unity. $U_i$ is the total potential which contains both $V_{\rm{LJ}}$ and $V_{\rm{FENE}}$.
In Eq.~(\ref{langevin}),  $T$ represents the quench temperature, measured in units of $\epsilon/k_B$. We set the value of $T$ well below the coil-globule transition temperature of a passive polymer to ensure  a globular conformation as the final steady state. Finally $R(t)$ stands for Gaussian noise with zero mean and unit variance. This is also  delta correlated over space and time, which can be represented as
\begin{equation}
\langle R_{i\mu}(t)R_{j\nu}(t')\rangle = \delta_{ij}\delta_{\mu \nu} \delta(t,t'),
\end{equation}
where $i,j$ represent the particle indices and $\mu$, $\nu$ correspond to the Cartesian coordinates. $\delta$ is the well-known Kronecker delta. The time step of integration $\delta t$ is chosen as $5 \times 10^{-4}$ in units of $\tau_0$, where $\tau_0 = \sqrt{m \sigma^2/\epsilon}$ is the unit of time. Determination of $\vec{r}_i, \dot{\vec{r}}_i$ for all the beads from  Eq.~(\ref{langevin}) with time provides the evolution of the passive polymer.

\par
~Then the activity for the beads is introduced in the Vicsek-like manner following the method described below \cite{vicsek, das1, skdas,paul2}.
After each MD step, the passive velocity for the $i$-th bead ($\vec{v}_i^{\rm{pas}}(t+\delta t)$) is modified by the active force ($\vec{f}_i$) which is defined as
\begin{equation}\label{active_force}
\vec{f}_i = f_A \hat{v}_i^{\rm{avg}} ,
\end{equation}
where $f_A$ measures the strength of activity. $f_A=0$ represents the case of the passive polymer. 
 $\hat{v}_i^{\rm{avg}}$ is the unit vector pointing in the average  direction of the velocities of all the beads within a spherical region of radius $r_v$ around the bead $i$. To calculate this we choose $r_v=r_c=2.5\sigma$. Then the passive velocity is modified as
\begin{equation}\label{resultant}
\vec{v}_i^{{*}}(t+\delta t) = \vec{v}_i^{\rm{pas}}(t+\delta t) + \frac{\vec{f}_i}{m_i} \delta t,
\end{equation}
by the implication of the active force. Thus the active force would change both the direction and magnitude of the velocity. The change in magnitude may increase the temperature of the 
system, which is not desired. Thus to keep the temperature of the polymer to the quenching value we rescale the magnitude of $\vec{v}_i^*$ to its passive value. This is done by
\begin{equation}
\vec{v}_i^{{f}}(t+\delta t) = \mid \vec{v}_i^{\rm{pas}}(t+\delta t) \mid\hat{n}_i,
\end{equation}
where $\hat{n}_i$ is the direction vector of $\vec{v}_i^*$. This procedure makes sure that  the application of Vicsek-like activity only changes the direction of the velocities without altering their magnitude.
Increase of the strength of the active force $\vec{f}_i$, by varying $f_A$,  will help the velocities of the beads to align themselves more rapidly. 
\par
~The initial configurations have been prepared at high temperature or good solvent condition where the conformation of the polymer is an extended coil.  These extended coil polymers were then quenched
 to a temperature $T=0.5$, well below the coil-globule transition temperature ($T_{\theta}$) for the passive case \cite{majumder3}. The results presented in the paper are for polymer chains with $N=256$ and $512$, where $N$ is the number of beads in it, the length of the polymer. In  each of the cases, all presented data have been averaged  over $100$ independent initial realizations.\\

\section{ Results}
~Before looking into the microscopic details  of the final \textit{pseudo equilibrium}  conformations of the polymer first we will look at a few quantities during its kinetics from coil to the globule conformation. The pathway for such transitions is quite complex, details of which will be presented elsewhere. Here we will focus on the quantities that are most relevant for the following discussion of the \textit{pseudo equilibrium} steady state conformations.
\par
\begin{figure}[t!]
	\centering
	\includegraphics*[width=0.46\textwidth]{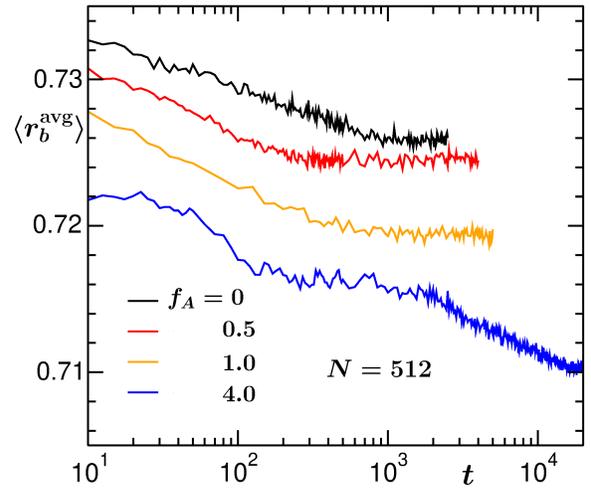}
	\caption{\label{rb_time} Semi-log plots of the average bond distance versus time for different values of $f_A$ for $N=512$. $r_b^{\rm{avg}}$ for each time was calculated from the first moment of $P(r_b)$. Here $\langle ..\rangle$ denotes the average over different initial conformations.}
\end{figure}

 
In Fig.~\ref{rb_time} we plot the average bond length $\langle r_b^{\rm{avg}}\rangle$ versus $t$ for different values of $f_A$. The bond length corresponding to any two consecutive beads, say, $i$ and $i+1$, is defined as
\begin{equation}
r_b=|\vec{r}_{i+1}-\vec{r}_i|,
\end{equation}
where $\vec{r}_i$ denotes the position of the $i$-th bead. Then the average bond length at each time can be calculated from the first moment of the corresponding distribution function as
\begin{equation}\label{bond_avg}
\langle r_b^{\rm{avg}}\rangle = \int r_b P(r_b,t) dr_b \,.
\end{equation}
In Fig.~\ref{rb_time} $\langle...\rangle$ represents the average over different independent initial conformations. We see that the plateau value at which the mean bond distance $\langle r_b^{\rm{avg}} \rangle$ saturates decreases with the increase of $f_A$. The saturation of $\langle r_b^{\rm{avg}}\rangle$ will help us to identify the onset of the steady state conformations of the polymer for further analyses.
\begin{figure}[t!]
	\centering
	\includegraphics*[width=0.46\textwidth]{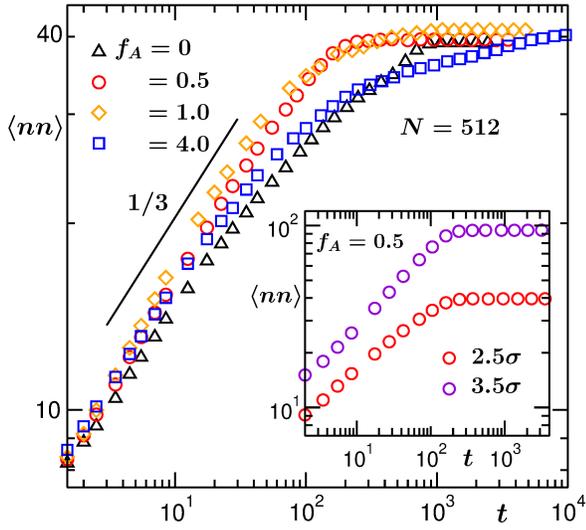}
	\caption{\label{nn_time} Log-log plots of the variation of average nearest neighbors ($nn$)  versus time for different values of $f_A$ for $N=512$ and $r_n=r_c=2.5\sigma$. Here also, $\langle ... \rangle$  indicates the averaging over different initial conformations. The solid line there represents a power law with exponent $1/3$. Inset shows the log-log plot of $nn$ versus $t$ for $f_A=0.5$ with two different choices of $r_n$ mentioned in the figure.}
\end{figure}
\par 
As already mentioned, in this paper we are interested in the structure and motion of the polymer at its \textit{pseudo equilibrium}. As there is always an attractive force among the monomers, this will help the beads to come closer and form a single cluster. It is expected that as we move forward in time during the evolution the average coordination number (nearest-neighbor beads) for a monomer increases. The number of nearest neighbors ($nn$) is calculated by counting the number of beads around any bead within a sphere of radius $r_n=r_c=2.5\sigma$. If $nn$ saturates to some value, then the time corresponding to the beginning of this saturation will denote the onset of the globular state. To check for that in Fig.~\ref{nn_time} we plot $\langle nn\rangle$, averaged over all the monomers and different initial conformations, versus $t$ for all the $f_A$  values as considered in the previous figure. We see that initially it increases more rapidly following a power-law behavior, $\langle nn \rangle \sim t^{1/3}$ until $\langle nn\rangle$ saturates towards the same value $\sim 40$ for all activity strengths $f_A$. But the times, say $t_n^s$, at which $\langle nn\rangle$ reach there are different for different values of $f_A$. It is obvious to visualize that if the conformation is a globular one then different choices of $r_n$ should lead to different values of the saturation of $\langle nn \rangle$. This we have shown in the inset of Fig.~\ref{nn_time} only for $f_A=0.5$. There we plot $\langle nn \rangle$  versus $t$ for two values of $r_n$, i.e., $2.5\sigma$ and $3.5\sigma$. Indeed the saturation value is much higher ($\sim 100$) for $r_n=3.5\sigma$. Also it seems like the exponent for the initial power-law growth of $\langle nn \rangle$ is higher for the larger choice of $r_n$. Most importantly the saturation time $t_n^s$ is independent of the choice of $r_n$. This feature is similar for the other values of $f_A$ as well. Now coming back to the main figure, we see a non-monotonic  behavior for $t_n^s$. For $f_A=0.5$ and $1.0$, the corresponding times are smaller than for the passive case, whereas for $f_A=4.0$, the value of $t_n^s$ is much higher. This fact is quite interesting and also demands for  further detailed analysis of the nonequilibrium kinetics of the globule formation.
\begin{figure}[t!]
	\centering
	\includegraphics*[width=0.48\textwidth]{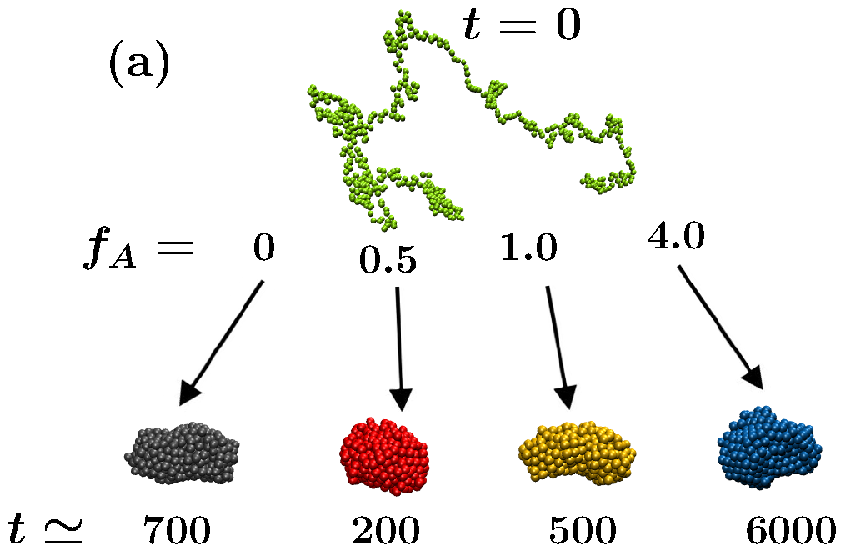}
	\vskip 0.4cm
	\includegraphics*[width=0.4\textwidth]{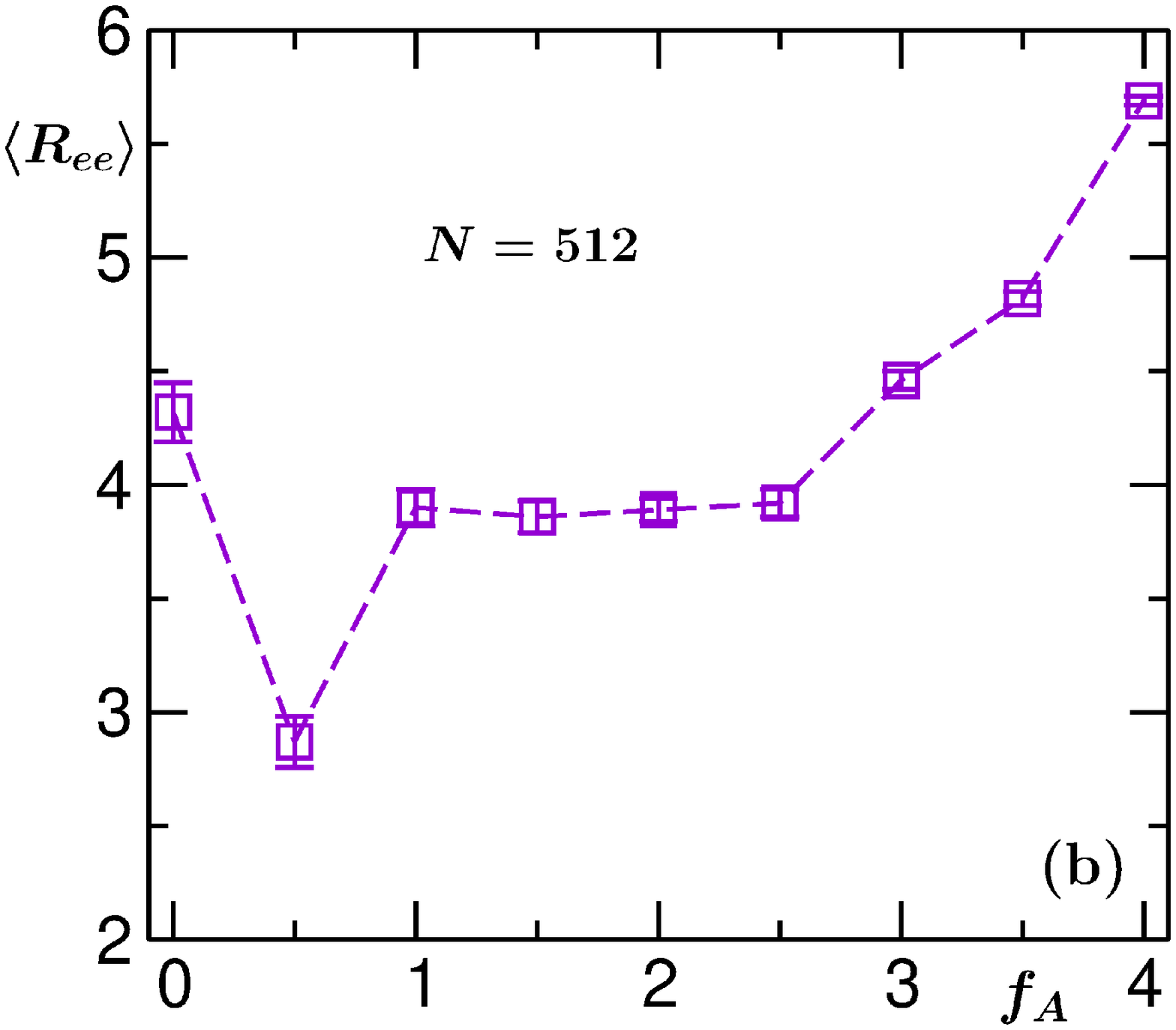}
	\caption{\label{snap} (a) Snapshots showing the {\em pseudo equilibrium} globular conformations with $N=512$ for the passive ($f_A=0$) as well as the active cases with $f_A=0.5$, $1.0$ and $4.0$. For all cases the starting conformation is the same which is shown with $t=0$. The corresponding times, mentioned below each of the globules, are the times at which a single cluster forms. (b) Plot of the average end-to-end distance $\langle R_{ee} \rangle$ for the globular conformations of the polymer versus $f_A$.}
\end{figure}
\par 
The preceding discussions were related to the nonequilibrium kinetics of the polymer which helped us to understand how and when the steady state has been reached. Next we focus on the main subject of our paper. First in Fig.~\ref{snap}(a) we show the 
\textit{pseudo equilibrium} conformations for the passive as well as for the active cases. In all the cases we started with a coil state of the polymer (for which  $t=0$ is mentioned) and see that the final conformations of the polymer are the globules. The times mentioned below each of these conformations correspond to the times ($t_s^g$) at which the globule forms. The corresponding times here were picked from the starting value of the saturation of $\langle nn \rangle$ shown in Fig.~\ref{nn_time}. Also this fact was confirmed by counting the number of clusters formed along the chain. Thus $t_s^g$ corresponds to the time when the number of clusters along the chain becomes $1$. After that there will be final rearrangements of the beads within this cluster to form a compact structure in order to minimize the surface energy \cite{schnabel}. Thus the saturation of $\langle r_b^{\rm{avg}}\rangle$ in Fig.~\ref{rb_time} occurs little later than for $\langle nn \rangle$. But one should note here that once a globule forms it is not possible to break it, as there is always an attractive force among the non-bonded monomers. Note that a completely repulsive potential, along with the Vicsek-like active force, is not suitable to produce a globular conformation of the polymer. We have explicitly checked this fact by using the WCA potential (\ref{wca_pot}) with different values of $f_A$ and chain lengths $N$ varying between $32$ and $128$.

\par 
Though the final conformations are qualitatively  similar in all the cases,  now we want to look  whether there exists any microscopic structural differences for different values of $f_A$.  In this regard, measurements of the end-to-end distance ($R_{ee}$) can give an idea of the spatial extension of the polymer in its globular conformation. $R_{ee}$, for a polymer, is calculated as 
	\begin{equation}
		R_{ee} = |{\vec{r}_1-\vec{r}_{N}}|,
	\end{equation}
	where $\vec{r}_1$ and $\vec{r}_N$ are the positions of the first and  last bead, respectively. In Fig.~\ref{snap}(b), we plot $\langle R_{ee}\rangle$ versus $f_A$. $R_{ee}$ has been averaged over different \textit{pseudo equilibrium} conformations.  There we observe a non-monotonic behavior as a function of $f_A$. Initially $R_{ee}$ decreases and for $f_A =0.5$ it attains a lower value than in the passive case indicating formation of a more compact globule. Then with the increase of $f_A$ we see that $R_{ee}$ again increases, and with much higher values of $f_A (\ge 3.5)$ it exceeds the value corresponding to the passive case. This points towards a deviation from spherical shape and formation of slightly elongated conformations with increasing activity.
\par
\begin{figure}[t!]
	\centering
	\includegraphics*[width=0.46\textwidth]{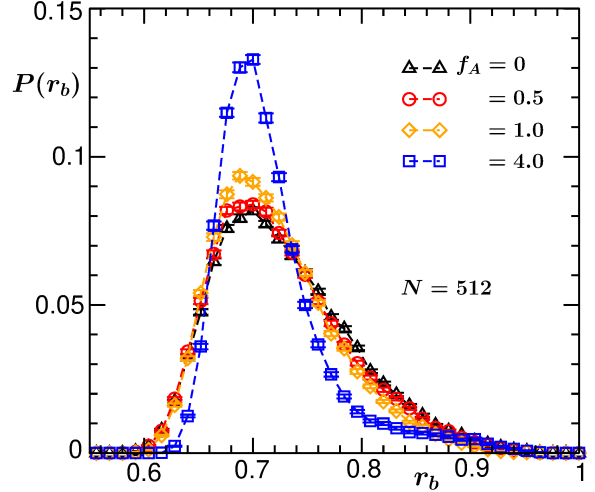}
	\caption{\label{rb_prob} Probability distribution of the bond lengths of the \textit{pseudo equilibrium} conformations of the polymer for the passive as well as for the active cases for the chain length $N=512$.}
\end{figure}

 
  From Fig.~\ref{rb_time} we already got a hint that the average bond distance decreases with $f_A$. Now in Fig.~\ref{rb_prob} we plot the distribution (normalized) of the bond distances for the passive as well as for the active cases in the steady state. It appears that in all the cases the distributions are non-Gaussian. Also it can be observed that with the increase of the strength of the activity, the peak height of the distribution increases and its width (a measure of the variance, the second moment of the distribution) decreases. We checked that for $f_A=4.0$ the width of the distribution is $\sim 55\%$ compared to that for the passive case. For all of them we see that the distributions are asymmetric with respect to their corresponding mean and have positive skewness which decreases with the increase of $f_A$. This fact indicates that when the activity overcomes the thermal noise, fluctuations  in the bond distances decrease.  From these  plots of the distributions it is hard to visualize whether there is any shift of the peak position in the abscissa variable. The position of the peak is essentially a measure for the average bond distance, which, as already observed from Fig.~\ref{rb_time}, decreases with the increase of $f_A$. Such changes appear in the third decimal place and are not easily identifiable from Fig.~\ref{rb_prob}. As the velocities of all the beads are aligned in a particular direction, thermal fluctuations play less role in determining the values of $r_b$. Thus  for the active case a more directed trajectory than for the passive polymer should be expected. 

\par
\begin{figure}[t!]
	\centering
	\includegraphics*[width=0.44\textwidth]{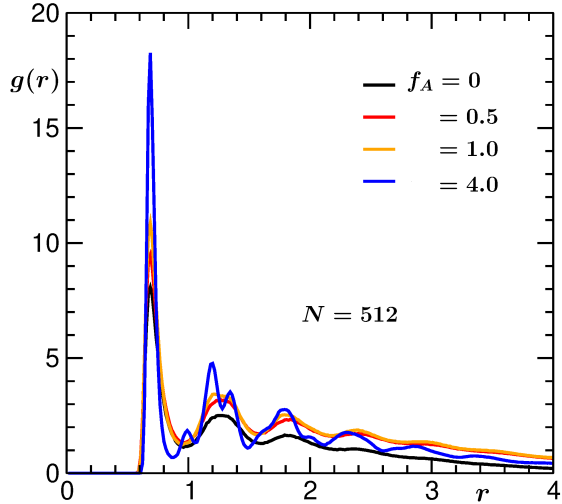}
	\caption{\label{g_r} Plots of the radial distribution functions $g(r)$ versus $r$ which measures the distance from a bead, for the passive as well as for the active cases for $N=512$.}
\end{figure}

After looking at the microscopic effect of the active force on the bond distances, now we  look whether there are structural differences in the conformations of the polymer in its steady state condition, for which a good candidate is the calculation of the radial distribution function. 
The radial distribution function $g(r)$, a measure for the average local density around a monomer, is calculated as
\begin{equation}\label{radial_distri}
	g(r)=\frac{n(r)}{4\pi r^2 \delta r},
\end{equation}
where $n(r)$ represents the average number of monomers around a bead within a shell of radius $r$ and thickness $\delta r$. In Fig.~\ref{g_r} we plot $g(r)$ versus $r$ for the passive as well as for the active cases for the steady state. From this plot we see that the positions of the first peak are at nearly the same value of $r$, which is equal to $2^{1/6}\sigma$, for all the cases. But their heights increase with activity. The positions and heights of the subsequent peaks for $f_A=0.5$ are more or less the same as for the passive case. But for the higher activities they differ from the $f_A=0$ case. We see that with further increase of activity the positions of the peaks (second, third, etc.) shift towards left and their heights increase, depicting the increase of the local density. Shifting of the peak positions towards left with the increase of activity suggests the lowering of the average bond length, which was also observed from Fig.~\ref{rb_time}.
\begin{figure}[t!]
	\centering
	\includegraphics*[width=0.46\textwidth]{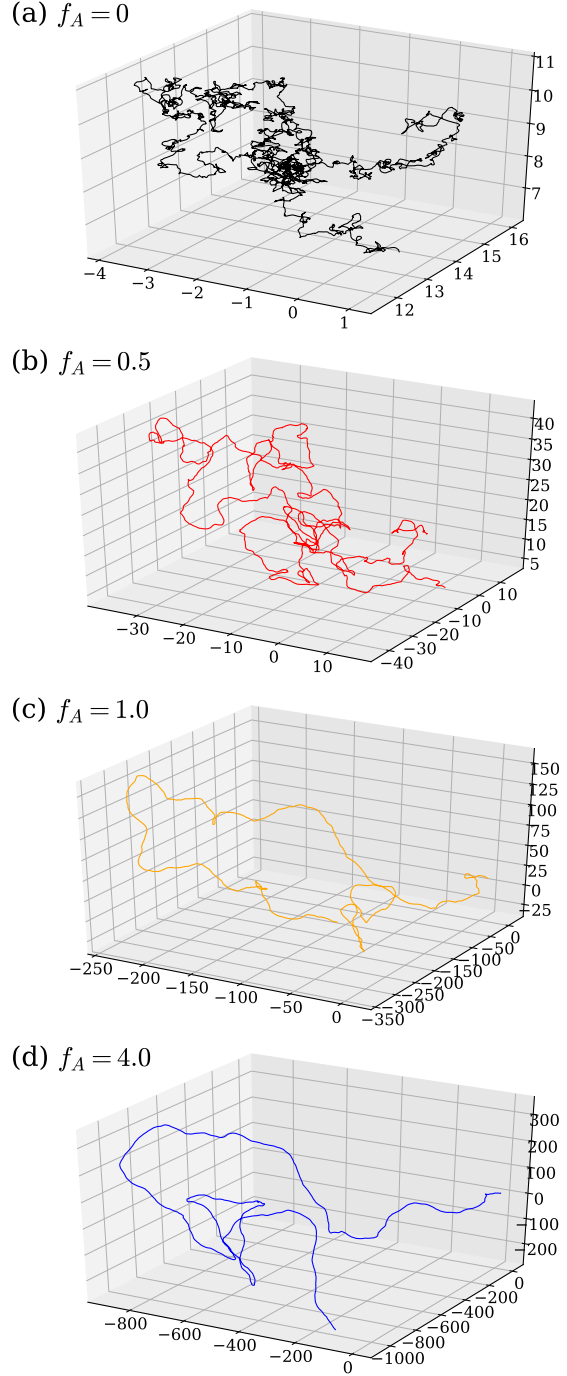}
	\caption{\label{trajec} Plots of the trajectories of the center-of-mass of the polymer for different values of $f_A$ for $N=256$ over a period of $400 \tau_0$ in the steady state. Different axis ticks in all the plots denote the $x$, $y$ and $z$ coordinates.}
\end{figure}
\begin{figure}[t!]
	\centering
	\includegraphics*[width=0.48\textwidth]{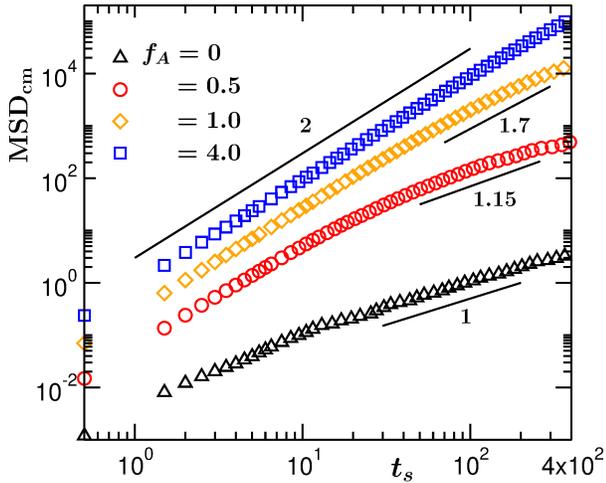}
	\caption{\label{msd_com} Log-log plot of the mean squared displacements of the center-of-mass of the polymer ($\rm{MSD}_{cm}$) versus time for different values of $f_A$. Here the time on the abscissa is denoted by $t_s ~(= t-t_0)$ which measures the translated time. The different solid lines show power laws with exponents mentioned next to them. All the results are for $N=256$.}
\end{figure}
\par 
 For our implementation of activity it is expected that as $f_A$ increases the velocities of the beads will be stronger aligned with each other.
	Thus we want to directly quantify how the Vicsek-like alignment activity has an effect on the motion of the polymer.  This has been done by tracking the motion of the center-of-mass of the polymer as well as a tagged monomer in the steady state. The center-of-mass of the polymer is defined as 
\begin{equation}\label{com_def}
\vec{r}_{\rm{cm}}=\frac{1}{N}\sum_{i=1}^{N} \vec{r}_i,
\end{equation}
where $\vec{r}_i$ is the position of the $i$-th bead. In Figs.~\ref{trajec}(a)-(d)  we plot the corresponding trajectories of $\vec{r}_{\rm{cm}}(t)$ for the passive as well as for the active cases during its time evolution in the steady state. For the passive polymer the trajectory follows a Brownian motion. As expected the motion of the polymer becomes more directed with the increase of $f_A$. For the active cases the polymer travels over a longer distance than in  the passive case. This fact can be appreciated by looking at the ranges of the $x$, $y$ and $z$ axes for all the cases. We also decided to look at the behavior of a tagged monomer \cite{chaki,shi2018,milchev_94}. For our analysis, without loss of generality, we considered the central bead. The trajectories of a tagged monomer show a similar trend, i.e., more directed motion, with increasing activity.

\par 
Now to look at the behavior of the motion at a quantitative level, we calculate the mean squared displacement ($\rm{MSD}$) of the center-of-mass of the polymer as well as of a tagged monomer. {The MSD for any object is defined as
	\begin{equation}
	{\rm{MSD}} = \langle [\vec{r}(t)-\vec{r}(t_0)]^2\rangle,
	\end{equation}
	where $\vec{r}(t)$ is the position of the object at time $t$ and $t_0$ represents the starting time of the measurement. Here, $\langle ...\rangle$ indicates averaging over different values of $t_0$ in the steady state trajectory. In general, MSD follows a power-law behavior in $t$,
\begin{equation}
	{\rm{MSD}} \sim t^{\alpha},
\end{equation}
where the exponent $\alpha =1$ corresponds to diffusive, $\alpha < 1$ to sub-diffusive and $\alpha > 1$ to super-diffusive motion, whereas  for ballistic motion one has $\alpha=2$.
\par 
In Fig.~\ref{msd_com} we plot the mean-squared-displacement of the center-of-mass $\rm{MSD}_{cm}$ versus $t_s=t-t_0$ for the values of $f_A$ as considered for Fig.~\ref{trajec}. Here $t_s$ defines the translated time, as it resets the time from the instant we start following the trajectory. In all the cases we see power-law behaviors with ${\rm{MSD}_{cm}} \sim t_s^{\alpha_{\rm{cm}}}$, where $\alpha_{\rm{cm}}$ is the corresponding exponent. For the passive polymer, we see an early regime, where the $\rm{MSD}_{cm}$ follows a ballistic-like behavior for a very short time followed by a crossover to the diffusive behavior. In these two regimes, $\rm{MSD}_{cm}$ follows power-law behaviors corresponding to $t_s^2$ and $t_s$, respectively. Now while increasing $f_A$, for $f_A=0.5$ and $1.0$, we see that the initial ballistic regime persists longer than  in the passive case and then it crosses over to super-diffusive behaviors with power-law exponents $> 1$. The corresponding exponents for these super-diffusive behaviors are mentioned in the figure adjacent to the data sets. To our understanding, even though the polymer model considered in Ref.~\cite{chaki} is different as to how activity is put in, a similar super-diffusive behavior for $\rm{MSD}_{cm}$ has been observed. Interestingly, for $f_A=4.0$ we see that the motion of the polymer becomes completely ballistic and the ${\rm{MSD}_{cm}} \sim t_s^2$  over the entire time range. We checked that with higher values of activity, the motion of the polymer remains ballistic but it travels over a longer distance within a particular time. Invoking analogy with a hard-sphere granular system where the particles move in a ballistic manner and align their velocities more parallel to each other upon inelastic collisions between them \cite{paul_epl}, here the polymer moves ballistically when the velocities of all the beads are perfectly aligned due to the implication of the active force.
\begin{figure}[t!]
	\centering
	\includegraphics*[width=0.48\textwidth]{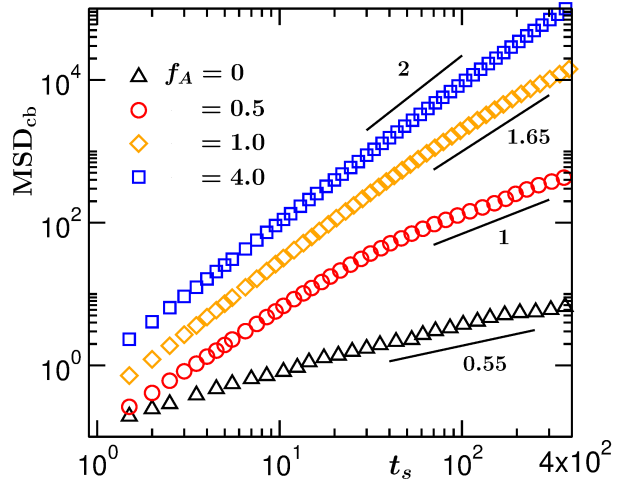}
	\caption{\label{msd_tag} Log-log plot of the mean squared displacements of the central monomer ($\rm{MSD}_{cb}$) versus translated time $t_s$ for different values of $f_A$. The solid lines represent power laws with exponents quoted just below them.}
\end{figure}
\par 
In the discussion above we have considered the polymer globule as a single entity by looking at  its center-of-mass motion. In its globular phase, it will  also be interesting to look at the behavior of a tagged monomer.  Here, in the globular conformation, any bead can be visualized as an active particle moving in a crowded environment created by the other beads surrounding it. For this, we looked at the $\rm{MSD}$ for the central bead of the chain. In Fig.~\ref{msd_tag} we plot $\rm{MSD}_{cb}$ versus the translated time $t_s$, for the same values of $f_A$ as in Fig.~\ref{msd_com}.
	For these, $\rm{MSD}_{cb}$ follows power-law behaviors with exponent $\alpha_{\rm{cb}}$ as ${\rm{MSD}_{cb}} \sim t_s^{\alpha_{\rm{cb}}}$.
 Interestingly, for the passive case we see that $\rm{MSD}_{cb}$ shows a sub-diffusive behavior with  $\alpha_{\rm{cb}} \simeq 0.55$ much smaller than the corresponding exponent $\alpha_{\rm{cm}} =1$. Similar anomalous diffusion for a tagged monomer has been observed earlier also for the collapsed conformation of a polymer chain \cite{milchev_94}.  Now for the lower activity, i.e., with $f_A=0.5$, it shows a diffusive motion with $\alpha_{\rm{cb}} \simeq 1.0$ although $\rm{MSD}_{cm}$ shows super-diffusive behavior with $\alpha_{\rm{cm}} \simeq 1.15$. For $f_A=1.0$, the motion of $\rm{MSD}_{cb}$ becomes super-diffusive with $\alpha_{\rm{cb}} \simeq 1.65$, comparable to the corresponding exponent for the $\rm{MSD}_{cm}$ ($\alpha_{\rm{cm}} \simeq 1.7$). For a much higher activity (i.e., with $f_A=4.0$) we see $\rm{MSD}_{cb}$ shows a ballistic behavior with $\alpha_{\rm{cb}}=2$ same as $\alpha_{\rm{cm}}$. Thus, with increasing activity, we see that difference between the exponents $\alpha_{\rm{cm}}$ and $\alpha_{\rm{cb}}$ decreases. With higher Vicsek-like activities when activity dominates over the thermal noise, the dynamics is controlled by the former.  Then in the steady state all beads move coherently in a particular direction. Thus the behavior of any tagged monomer becomes very much similar to that of the center-of-mass of the polymer globule and any dissimilarity between the corresponding exponents disappears.\\

\section{Conclusion}
~~We have studied the effect of Vicsek-like activity on the \textit{pseudo equilibrium} conformations and dynamics of a flexible homopolymer chain undergoing a coil-globule transition. 
To ensure that the temperature remains at our chosen value which is well below the collapse transition temperature for the passive polymer, a Langevin thermostat  has been employed during the MD simulation. Due to the active force 
the velocities of the beads align in a particular direction  decided by its neighbors. Whereas for the passive polymer the dynamics is mainly governed by the force due to thermal fluctuations acting on each bead, for the active case there is always a competition between this random and the active force.
\par 
~~The microscopic details of the structures were calculated by looking at the average bond length as well as its distribution in the globular conformation. We see that the fluctuations in the bond lengths as well as the average value decrease with the increase of activity. This has been confirmed via the calculation of  bond length distribution and the pair correlation function which is more relevant for an experimental measure. As the effect of activity is related to the velocity  alignment of the beads, the polymer with activity shows a more directed motion than its passive limit and can travel a much longer distance within a medium. To check for it at the quantitative level, in the globular phase of the polymer, we looked at the mean-squared-displacement of its center-of-mass as well as for a tagged monomer. For the passive polymer, its center-of-mass shows a diffusive motion, whereas the motion of the tagged central bead is sub-diffusive.  Interestingly, in both cases, this behavior changes with increasing active force over super-diffusive to ballistic motion. With higher activity, when all the beads are aligned perfectly in a certain direction, the motion of the center-of-mass becomes very much coherent with that of a tagged monomer. In our model, we have considered the solvent effects implicitly using the  parameter $\gamma$. Thus tuning the value of $\gamma$ along with the active force can give us more control over the motion of the polymer in its overdamped limit. In these regard, the nonequilibrium kinetics of globule formation will also be insightful.  Also it can be interesting to look at aging properties for its nonequilibrium kinetics. These questions we plan to tackle  in the future.\\
\par   
~~{\bf Acknowledgement:} 
This project was funded by the Deutsche Forschungsgemeinschaft (DFG, German Research Foundation) 
under Grant No.\ 189\,853\,844--SFB/TRR 102 (Project B04). It was further supported by the Deutsch-Franz\"osische Hochschule (DFH-UFA) 
through the Doctoral College ``$\mathbb{L}^4$'' under Grant No.\ CDFA-02-07, the Leipzig Graduate School of Natural Sciences ``BuildMoNa'', 
and the EU COST programme EUTOPIA under Grant No.\ CA17139.


\begin{thebibliography}{46}%
	\makeatletter
	\providecommand \@ifxundefined [1]{%
		\@ifx{#1\undefined}
	}%
	\providecommand \@ifnum [1]{%
		\ifnum #1\expandafter \@firstoftwo
		\else \expandafter \@secondoftwo
		\fi
	}%
	\providecommand \@ifx [1]{%
		\ifx #1\expandafter \@firstoftwo
		\else \expandafter \@secondoftwo
		\fi
	}%
	\providecommand \natexlab [1]{#1}%
	\providecommand \enquote  [1]{``#1''}%
	\providecommand \bibnamefont  [1]{#1}%
	\providecommand \bibfnamefont [1]{#1}%
	\providecommand \citenamefont [1]{#1}%
	\providecommand \href@noop [0]{\@secondoftwo}%
	\providecommand \href [0]{\begingroup \@sanitize@url \@href}%
	\providecommand \@href[1]{\@@startlink{#1}\@@href}%
	\providecommand \@@href[1]{\endgroup#1\@@endlink}%
	\providecommand \@sanitize@url [0]{\catcode `\\12\catcode `\$12\catcode
		`\&12\catcode `\#12\catcode `\^12\catcode `\_12\catcode `\%12\relax}%
	\providecommand \@@startlink[1]{}%
	\providecommand \@@endlink[0]{}%
	\providecommand \url  [0]{\begingroup\@sanitize@url \@url }%
	\providecommand \@url [1]{\endgroup\@href {#1}{\urlprefix }}%
	\providecommand \urlprefix  [0]{URL }%
	\providecommand \Eprint [0]{\href }%
	\providecommand \doibase [0]{http://dx.doi.org/}%
	\providecommand \selectlanguage [0]{\@gobble}%
	\providecommand \bibinfo  [0]{\@secondoftwo}%
	\providecommand \bibfield  [0]{\@secondoftwo}%
	\providecommand \translation [1]{[#1]}%
	\providecommand \BibitemOpen [0]{}%
	\providecommand \bibitemStop [0]{}%
	\providecommand \bibitemNoStop [0]{.\EOS\space}%
	\providecommand \EOS [0]{\spacefactor3000\relax}%
	\providecommand \BibitemShut  [1]{\csname bibitem#1\endcsname}%
	\let\auto@bib@innerbib\@empty
	\bibitem [{\citenamefont {Ramaswamy}(2010)}]{ramasw1}%
	\BibitemOpen
	\bibfield  {author} {\bibinfo {author} {\bibfnamefont {S.}~\bibnamefont
			{Ramaswamy}},\ }\bibfield  {title} {\enquote {\bibinfo {title} {The mechanics
				and statistics of active matter},}\ }\href {\doibase
		10.1146/annurev-conmatphys-070909-104101} {\bibfield  {journal} {\bibinfo
			{journal} {Ann. Rev. Cond. Mat. Phys.}\ }\textbf {\bibinfo {volume} {1}},\
		\bibinfo {pages} {323--345} (\bibinfo {year} {2010})}\BibitemShut {NoStop}%
	\bibitem [{\citenamefont {Romanczuk}\ \emph {et~al.}(2012)\citenamefont
		{Romanczuk}, \citenamefont {B{\"a}r}, \citenamefont {Ebeling}, \citenamefont
		{Lindner},\ and\ \citenamefont {Schimansky}}]{roman}%
	\BibitemOpen
	\bibfield  {author} {\bibinfo {author} {\bibfnamefont {P.}~\bibnamefont
			{Romanczuk}}, \bibinfo {author} {\bibfnamefont {M.}~\bibnamefont {B{\"a}r}},
		\bibinfo {author} {\bibfnamefont {W.}~\bibnamefont {Ebeling}}, \bibinfo
		{author} {\bibfnamefont {B.}~\bibnamefont {Lindner}}, \ and\ \bibinfo
		{author} {\bibfnamefont {L~.-G.}\ \bibnamefont {Schimansky}},\ }\bibfield
	{title} {\enquote {\bibinfo {title} {Active {B}rownian particles},}\
	}\href@noop {} {\bibfield  {journal} {\bibinfo  {journal} {Eur. Phys. J.
				Spec. Top.}\ }\textbf {\bibinfo {volume} {202}},\ \bibinfo {pages} {1--162}
		(\bibinfo {year} {2012})}\BibitemShut {NoStop}%
	\bibitem [{\citenamefont {Cates}\ and\ \citenamefont
		{Tailleur}(2015)}]{cates1}%
	\BibitemOpen
	\bibfield  {author} {\bibinfo {author} {\bibfnamefont {M.~E.}\ \bibnamefont
			{Cates}}\ and\ \bibinfo {author} {\bibfnamefont {J.}~\bibnamefont
			{Tailleur}},\ }\bibfield  {title} {\enquote {\bibinfo {title}
			{Motility-induced phase separation},}\ }\href {\doibase
		10.1146/annurev-conmatphys-031214-014710} {\bibfield  {journal} {\bibinfo
			{journal} {Ann. Rev. Cond. Mat. Phys.}\ }\textbf {\bibinfo {volume} {6}},\
		\bibinfo {pages} {219--244} (\bibinfo {year} {2015})}\BibitemShut {NoStop}%
	\bibitem [{\citenamefont {Elgeti}\ \emph {et~al.}(2015)\citenamefont {Elgeti},
		\citenamefont {Winkler},\ and\ \citenamefont {Gompper}}]{elgeti1}%
	\BibitemOpen
	\bibfield  {author} {\bibinfo {author} {\bibfnamefont {J.}~\bibnamefont
			{Elgeti}}, \bibinfo {author} {\bibfnamefont {R.~G.}\ \bibnamefont {Winkler}},
		\ and\ \bibinfo {author} {\bibfnamefont {G.}~\bibnamefont {Gompper}},\
	}\bibfield  {title} {\enquote {\bibinfo {title} {Physics of
				microswimmers{\textemdash}single particle motion and collective behavior: A
				review},}\ }\href {\doibase 10.1088/0034-4885/78/5/056601} {\bibfield
		{journal} {\bibinfo  {journal} {Rep. Prog. Phys.}\ }\textbf {\bibinfo
			{volume} {78}},\ \bibinfo {pages} {056601} (\bibinfo {year}
		{2015})}\BibitemShut {NoStop}%
	\bibitem [{\citenamefont {Winkler}\ and\ \citenamefont
		{Gompper}(2020)}]{winkler}%
	\BibitemOpen
	\bibfield  {author} {\bibinfo {author} {\bibfnamefont {R~G.}\ \bibnamefont
			{Winkler}}\ and\ \bibinfo {author} {\bibfnamefont {G.}~\bibnamefont
			{Gompper}},\ }\bibfield  {title} {\enquote {\bibinfo {title} {The physics of
				active polymers and filaments},}\ }\href {\doibase 10.1063/5.0011466}
	{\bibfield  {journal} {\bibinfo  {journal} {J. Chem. Phys.}\ }\textbf
		{\bibinfo {volume} {153}},\ \bibinfo {pages} {040901} (\bibinfo {year}
		{2020})}\BibitemShut {NoStop}%
	\bibitem [{\citenamefont {Shaebani}\ \emph {et~al.}(2020)\citenamefont
		{Shaebani}, \citenamefont {Wysocki}, \citenamefont {Winkler}, \citenamefont
		{Gompper},\ and\ \citenamefont {Rieger}}]{shaeb}%
	\BibitemOpen
	\bibfield  {author} {\bibinfo {author} {\bibfnamefont {M.R.}\ \bibnamefont
			{Shaebani}}, \bibinfo {author} {\bibfnamefont {A.}~\bibnamefont {Wysocki}},
		\bibinfo {author} {\bibfnamefont {R.G.}\ \bibnamefont {Winkler}}, \bibinfo
		{author} {\bibfnamefont {G.}~\bibnamefont {Gompper}}, \ and\ \bibinfo
		{author} {\bibfnamefont {H.}~\bibnamefont {Rieger}},\ }\bibfield  {title}
	{\enquote {\bibinfo {title} {Computational models for active matter},}\
	}\href@noop {} {\bibfield  {journal} {\bibinfo  {journal} {Nat. Rev. Phys.}\
		}\textbf {\bibinfo {volume} {2}},\ \bibinfo {pages} {181} (\bibinfo {year}
		{2020})}\BibitemShut {NoStop}%
	\bibitem [{\citenamefont {Vicsek}\ \emph {et~al.}(1995)\citenamefont {Vicsek},
		\citenamefont {Czir\'ok}, \citenamefont {Ben-Jacob}, \citenamefont {Cohen},\
		and\ \citenamefont {Shochet}}]{vicsek}%
	\BibitemOpen
	\bibfield  {author} {\bibinfo {author} {\bibfnamefont {T.}~\bibnamefont
			{Vicsek}}, \bibinfo {author} {\bibfnamefont {A.}~\bibnamefont {Czir\'ok}},
		\bibinfo {author} {\bibfnamefont {E.}~\bibnamefont {Ben-Jacob}}, \bibinfo
		{author} {\bibfnamefont {I.}~\bibnamefont {Cohen}}, \ and\ \bibinfo {author}
		{\bibfnamefont {O.}~\bibnamefont {Shochet}},\ }\bibfield  {title} {\enquote
		{\bibinfo {title} {Novel {T}ype of {P}hase {T}ransition in a {S}ystem of
				{S}elf-{D}riven {P}articles},}\ }\href {\doibase 10.1103/PhysRevLett.75.1226}
	{\bibfield  {journal} {\bibinfo  {journal} {Phys. Rev. Lett.}\ }\textbf
		{\bibinfo {volume} {75}},\ \bibinfo {pages} {1226--1229} (\bibinfo {year}
		{1995})}\BibitemShut {NoStop}%
	\bibitem [{\citenamefont {Tailleur}\ and\ \citenamefont
		{Cates}(2008)}]{tailleur}%
	\BibitemOpen
	\bibfield  {author} {\bibinfo {author} {\bibfnamefont {J.}~\bibnamefont
			{Tailleur}}\ and\ \bibinfo {author} {\bibfnamefont {M.~E.}\ \bibnamefont
			{Cates}},\ }\bibfield  {title} {\enquote {\bibinfo {title} {Statistical
				{M}echanics of {I}nteracting {R}un-and-{T}umble {B}acteria},}\ }\href
	{\doibase 10.1103/PhysRevLett.100.218103} {\bibfield  {journal} {\bibinfo
			{journal} {Phys. Rev. Lett.}\ }\textbf {\bibinfo {volume} {100}},\ \bibinfo
		{pages} {218103} (\bibinfo {year} {2008})}\BibitemShut {NoStop}%
	\bibitem [{\citenamefont {Chat\'{e}}\ \emph {et~al.}(2008)\citenamefont
		{Chat\'{e}}, \citenamefont {Ginelli}, \citenamefont {Gr\'{e}goire},
		\citenamefont {Peruani},\ and\ \citenamefont {Raynaud}}]{chate1}%
	\BibitemOpen
	\bibfield  {author} {\bibinfo {author} {\bibfnamefont {H.}~\bibnamefont
			{Chat\'{e}}}, \bibinfo {author} {\bibfnamefont {F.}~\bibnamefont {Ginelli}},
		\bibinfo {author} {\bibfnamefont {G.}~\bibnamefont {Gr\'{e}goire}}, \bibinfo
		{author} {\bibfnamefont {F.}~\bibnamefont {Peruani}}, \ and\ \bibinfo
		{author} {\bibfnamefont {F.}~\bibnamefont {Raynaud}},\ }\bibfield  {title}
	{\enquote {\bibinfo {title} {Modeling collective motion: {V}ariations on the
				{V}icsek model},}\ }\href {\doibase 10.1140/epjb/e2008-00275-9} {\bibfield
		{journal} {\bibinfo  {journal} {Eur. Phys. J. B}\ }\textbf {\bibinfo {volume}
			{64}},\ \bibinfo {pages} {451--456} (\bibinfo {year} {2008})}\BibitemShut
	{NoStop}%
	\bibitem [{\citenamefont {Mishra}\ \emph {et~al.}(2010)\citenamefont {Mishra},
		\citenamefont {Baskaran},\ and\ \citenamefont {Marchetti}}]{mishra2}%
	\BibitemOpen
	\bibfield  {author} {\bibinfo {author} {\bibfnamefont {S.}~\bibnamefont
			{Mishra}}, \bibinfo {author} {\bibfnamefont {A.}~\bibnamefont {Baskaran}}, \
		and\ \bibinfo {author} {\bibfnamefont {M.~C.}\ \bibnamefont {Marchetti}},\
	}\bibfield  {title} {\enquote {\bibinfo {title} {Fluctuations and pattern
				formation in self-propelled particles},}\ }\href {\doibase
		10.1103/PhysRevE.81.061916} {\bibfield  {journal} {\bibinfo  {journal} {Phys.
				Rev. E}\ }\textbf {\bibinfo {volume} {81}},\ \bibinfo {pages} {061916}
		(\bibinfo {year} {2010})}\BibitemShut {NoStop}%
	\bibitem [{\citenamefont {Loi}\ \emph {et~al.}(2011)\citenamefont {Loi},
		\citenamefont {Mossa},\ and\ \citenamefont {Cugliandolo}}]{loi}%
	\BibitemOpen
	\bibfield  {author} {\bibinfo {author} {\bibfnamefont {D.}~\bibnamefont
			{Loi}}, \bibinfo {author} {\bibfnamefont {S.}~\bibnamefont {Mossa}}, \ and\
		\bibinfo {author} {\bibfnamefont {L.~F.}\ \bibnamefont {Cugliandolo}},\
	}\bibfield  {title} {\enquote {\bibinfo {title} {Non-conservative forces and
				effective temperatures in active polymers},}\ }\href {\doibase
		10.1039/C1SM05819C} {\bibfield  {journal} {\bibinfo  {journal} {Soft Matter}\
		}\textbf {\bibinfo {volume} {7}},\ \bibinfo {pages} {10193--10209} (\bibinfo
		{year} {2011})}\BibitemShut {NoStop}%
	\bibitem [{\citenamefont {Vicsek}\ and\ \citenamefont {Zafeiris}(2012)}]{vic2}%
	\BibitemOpen
	\bibfield  {author} {\bibinfo {author} {\bibfnamefont {T.}~\bibnamefont
			{Vicsek}}\ and\ \bibinfo {author} {\bibfnamefont {A.}~\bibnamefont
			{Zafeiris}},\ }\bibfield  {title} {\enquote {\bibinfo {title} {Collective
				motion},}\ }\href {\doibase https://doi.org/10.1016/j.physrep.2012.03.004}
	{\bibfield  {journal} {\bibinfo  {journal} {Phys. Rep.}\ }\textbf {\bibinfo
			{volume} {517}},\ \bibinfo {pages} {71 -- 140} (\bibinfo {year}
		{2012})}\BibitemShut {NoStop}%
	\bibitem [{\citenamefont {Fily}\ and\ \citenamefont {Marchetti}(2012)}]{fily}%
	\BibitemOpen
	\bibfield  {author} {\bibinfo {author} {\bibfnamefont {Y.}~\bibnamefont
			{Fily}}\ and\ \bibinfo {author} {\bibfnamefont {M.~C.}\ \bibnamefont
			{Marchetti}},\ }\bibfield  {title} {\enquote {\bibinfo {title} {Athermal
				{P}hase {S}eparation of {S}elf-{P}ropelled {P}articles with {N}o
				{A}lignment},}\ }\href {\doibase 10.1103/PhysRevLett.108.235702} {\bibfield
		{journal} {\bibinfo  {journal} {Phys. Rev. Lett.}\ }\textbf {\bibinfo
			{volume} {108}},\ \bibinfo {pages} {235702} (\bibinfo {year}
		{2012})}\BibitemShut {NoStop}%
	\bibitem [{\citenamefont {Redner}\ \emph {et~al.}(2013)\citenamefont {Redner},
		\citenamefont {Hagan},\ and\ \citenamefont {Baskaran}}]{redner}%
	\BibitemOpen
	\bibfield  {author} {\bibinfo {author} {\bibfnamefont {G.~S.}\ \bibnamefont
			{Redner}}, \bibinfo {author} {\bibfnamefont {M.~F.}\ \bibnamefont {Hagan}}, \
		and\ \bibinfo {author} {\bibfnamefont {A.}~\bibnamefont {Baskaran}},\
	}\bibfield  {title} {\enquote {\bibinfo {title} {Structure and {D}ynamics of
				a {P}hase-{S}eparating {A}ctive {C}olloidal {F}luid},}\ }\href {\doibase
		10.1103/PhysRevLett.110.055701} {\bibfield  {journal} {\bibinfo  {journal}
			{Phys. Rev. Lett.}\ }\textbf {\bibinfo {volume} {110}},\ \bibinfo {pages}
		{055701} (\bibinfo {year} {2013})}\BibitemShut {NoStop}%
	\bibitem [{\citenamefont {Stenhammar}\ \emph {et~al.}(2014)\citenamefont
		{Stenhammar}, \citenamefont {Marenduzzo}, \citenamefont {Allen},\ and\
		\citenamefont {Cates}}]{cates2}%
	\BibitemOpen
	\bibfield  {author} {\bibinfo {author} {\bibfnamefont {J.}~\bibnamefont
			{Stenhammar}}, \bibinfo {author} {\bibfnamefont {D.}~\bibnamefont
			{Marenduzzo}}, \bibinfo {author} {\bibfnamefont {R.~J.}\ \bibnamefont
			{Allen}}, \ and\ \bibinfo {author} {\bibfnamefont {M.~E.}\ \bibnamefont
			{Cates}},\ }\bibfield  {title} {\enquote {\bibinfo {title} {Phase behaviour
				of active {B}rownian particles: The role of dimensionality},}\ }\href
	{\doibase 10.1039/C3SM52813H} {\bibfield  {journal} {\bibinfo  {journal}
			{Soft Matter}\ }\textbf {\bibinfo {volume} {10}},\ \bibinfo {pages}
		{1489--1499} (\bibinfo {year} {2014})}\BibitemShut {NoStop}%
	\bibitem [{\citenamefont {Das}\ \emph {et~al.}(2014)\citenamefont {Das},
		\citenamefont {Egorov}, \citenamefont {Trefz}, \citenamefont {Virnau},\ and\
		\citenamefont {Binder}}]{skdas}%
	\BibitemOpen
	\bibfield  {author} {\bibinfo {author} {\bibfnamefont {S.~K.}\ \bibnamefont
			{Das}}, \bibinfo {author} {\bibfnamefont {S.~A.}\ \bibnamefont {Egorov}},
		\bibinfo {author} {\bibfnamefont {B.}~\bibnamefont {Trefz}}, \bibinfo
		{author} {\bibfnamefont {P.}~\bibnamefont {Virnau}}, \ and\ \bibinfo {author}
		{\bibfnamefont {K.}~\bibnamefont {Binder}},\ }\bibfield  {title} {\enquote
		{\bibinfo {title} {Phase {B}ehavior of {A}ctive {S}wimmers in {D}epletants:
				{M}olecular {D}ynamics and {I}ntegral {E}quation {T}heory},}\ }\href
	{\doibase 10.1103/PhysRevLett.112.198301} {\bibfield  {journal} {\bibinfo
			{journal} {Phys. Rev. Lett.}\ }\textbf {\bibinfo {volume} {112}},\ \bibinfo
		{pages} {198301} (\bibinfo {year} {2014})}\BibitemShut {NoStop}%
	\bibitem [{\citenamefont {Das}(2017)}]{das1}%
	\BibitemOpen
	\bibfield  {author} {\bibinfo {author} {\bibfnamefont {S.~K.}\ \bibnamefont
			{Das}},\ }\bibfield  {title} {\enquote {\bibinfo {title} {Pattern, growth,
				and aging in aggregation kinetics of a {V}icsek-like active matter model},}\
	}\href {\doibase 10.1063/1.4974256} {\bibfield  {journal} {\bibinfo
			{journal} {J. Chem. Phys.}\ }\textbf {\bibinfo {volume} {146}},\ \bibinfo
		{pages} {044902} (\bibinfo {year} {2017})}\BibitemShut {NoStop}%
	\bibitem [{\citenamefont {Paul}\ \emph {et~al.}(2021)\citenamefont {Paul},
		\citenamefont {Bera},\ and\ \citenamefont {Das}}]{paul1}%
	\BibitemOpen
	\bibfield  {author} {\bibinfo {author} {\bibfnamefont {S.}~\bibnamefont
			{Paul}}, \bibinfo {author} {\bibfnamefont {A.}~\bibnamefont {Bera}}, \ and\
		\bibinfo {author} {\bibfnamefont {S.~K.}\ \bibnamefont {Das}},\ }\bibfield
	{title} {\enquote {\bibinfo {title} {How do clusters in phase-separating
				active matter systems grow? {A} study for {V}icsek activity in systems
				undergoing vapor–solid transition},}\ }\href {\doibase 10.1039/D0SM01762K}
	{\bibfield  {journal} {\bibinfo  {journal} {Soft Matter}\ }\textbf {\bibinfo
			{volume} {17}},\ \bibinfo {pages} {645--654} (\bibinfo {year}
		{2021})}\BibitemShut {NoStop}%
	\bibitem [{\citenamefont {Jiang}\ \emph {et~al.}(2010)\citenamefont {Jiang},
		\citenamefont {Yoshinaga},\ and\ \citenamefont {Sano}}]{jiang}%
	\BibitemOpen
	\bibfield  {author} {\bibinfo {author} {\bibfnamefont {H.-R.}\ \bibnamefont
			{Jiang}}, \bibinfo {author} {\bibfnamefont {N.}~\bibnamefont {Yoshinaga}}, \
		and\ \bibinfo {author} {\bibfnamefont {M.}~\bibnamefont {Sano}},\ }\bibfield
	{title} {\enquote {\bibinfo {title} {Active {M}otion of a {J}anus {P}article
				by {S}elf-{T}hermophoresis in a {D}efocused {L}aser {B}eam},}\ }\href
	{\doibase 10.1103/PhysRevLett.105.268302} {\bibfield  {journal} {\bibinfo
			{journal} {Phys. Rev. Lett.}\ }\textbf {\bibinfo {volume} {105}},\ \bibinfo
		{pages} {268302} (\bibinfo {year} {2010})}\BibitemShut {NoStop}%
	\bibitem [{\citenamefont {Biswas}\ \emph {et~al.}(2017)\citenamefont {Biswas},
		\citenamefont {Manna}, \citenamefont {Laskar}, \citenamefont {Kumar},
		\citenamefont {Adhikari},\ and\ \citenamefont {Kumaraswamy}}]{biswas}%
	\BibitemOpen
	\bibfield  {author} {\bibinfo {author} {\bibfnamefont {B.}~\bibnamefont
			{Biswas}}, \bibinfo {author} {\bibfnamefont {R.~K.}\ \bibnamefont {Manna}},
		\bibinfo {author} {\bibfnamefont {A.}~\bibnamefont {Laskar}}, \bibinfo
		{author} {\bibfnamefont {P.~B.~S.}\ \bibnamefont {Kumar}}, \bibinfo {author}
		{\bibfnamefont {R.}~\bibnamefont {Adhikari}}, \ and\ \bibinfo {author}
		{\bibfnamefont {G.}~\bibnamefont {Kumaraswamy}},\ }\bibfield  {title}
	{\enquote {\bibinfo {title} {Linking catalyst-coated isotropic colloids into
				“active” flexible chains enhances their diffusivity},}\ }\href {\doibase
		10.1021/acsnano.7b04265} {\bibfield  {journal} {\bibinfo  {journal} {ACS
				Nano}\ }\textbf {\bibinfo {volume} {11}},\ \bibinfo {pages} {10025--10031}
		(\bibinfo {year} {2017})}\BibitemShut {NoStop}%
	\bibitem [{\citenamefont {I.-Holder}\ \emph {et~al.}(2015)\citenamefont
		{I.-Holder}, \citenamefont {Elgeti},\ and\ \citenamefont {Gompper}}]{holder}%
	\BibitemOpen
	\bibfield  {author} {\bibinfo {author} {\bibfnamefont {R.~E.}\ \bibnamefont
			{I.-Holder}}, \bibinfo {author} {\bibfnamefont {J.}~\bibnamefont {Elgeti}}, \
		and\ \bibinfo {author} {\bibfnamefont {G.}~\bibnamefont {Gompper}},\
	}\bibfield  {title} {\enquote {\bibinfo {title} {Self-propelled worm-like
				filaments: {S}pontaneous spiral formation{,} structure{,} and dynamics},}\
	}\href {\doibase 10.1039/C5SM01683E} {\bibfield  {journal} {\bibinfo
			{journal} {Soft Matter}\ }\textbf {\bibinfo {volume} {11}},\ \bibinfo {pages}
		{7181--7190} (\bibinfo {year} {2015})}\BibitemShut {NoStop}%
	\bibitem [{\citenamefont {Kaiser}\ \emph {et~al.}(2015)\citenamefont {Kaiser},
		\citenamefont {Babel}, \citenamefont {ten Hagen}, \citenamefont {von
			Ferber},\ and\ \citenamefont {L\"{o}wen}}]{kaiser}%
	\BibitemOpen
	\bibfield  {author} {\bibinfo {author} {\bibfnamefont {A.}~\bibnamefont
			{Kaiser}}, \bibinfo {author} {\bibfnamefont {S.}~\bibnamefont {Babel}},
		\bibinfo {author} {\bibfnamefont {B.}~\bibnamefont {ten Hagen}}, \bibinfo
		{author} {\bibfnamefont {C.}~\bibnamefont {von Ferber}}, \ and\ \bibinfo
		{author} {\bibfnamefont {H.}~\bibnamefont {L\"{o}wen}},\ }\bibfield  {title}
	{\enquote {\bibinfo {title} {How does a flexible chain of active particles
				swell?}}\ }\href {\doibase 10.1063/1.4916134} {\bibfield  {journal} {\bibinfo
			{journal} {J. Chem. Phys.}\ }\textbf {\bibinfo {volume} {142}},\ \bibinfo
		{pages} {124905} (\bibinfo {year} {2015})}\BibitemShut {NoStop}%
	\bibitem [{\citenamefont {Duman}\ \emph {et~al.}(2018)\citenamefont {Duman},
		\citenamefont {I.-Holder}, \citenamefont {Elgeti},\ and\ \citenamefont
		{Gompper}}]{duman}%
	\BibitemOpen
	\bibfield  {author} {\bibinfo {author} {\bibfnamefont {{\"{O}}.}~\bibnamefont
			{Duman}}, \bibinfo {author} {\bibfnamefont {R.~E.}\ \bibnamefont
			{I.-Holder}}, \bibinfo {author} {\bibfnamefont {J.}~\bibnamefont {Elgeti}}, \
		and\ \bibinfo {author} {\bibfnamefont {G.}~\bibnamefont {Gompper}},\
	}\bibfield  {title} {\enquote {\bibinfo {title} {Collective dynamics of
				self-propelled semiflexible filaments},}\ }\href {\doibase
		10.1039/C8SM00282G} {\bibfield  {journal} {\bibinfo  {journal} {Soft Matter}\
		}\textbf {\bibinfo {volume} {14}},\ \bibinfo {pages} {4483--4494} (\bibinfo
		{year} {2018})}\BibitemShut {NoStop}%
	\bibitem [{\citenamefont {Bianco}\ \emph {et~al.}(2018)\citenamefont {Bianco},
		\citenamefont {Locatelli},\ and\ \citenamefont {Malgaretti}}]{bianco}%
	\BibitemOpen
	\bibfield  {author} {\bibinfo {author} {\bibfnamefont {V.}~\bibnamefont
			{Bianco}}, \bibinfo {author} {\bibfnamefont {E.}~\bibnamefont {Locatelli}}, \
		and\ \bibinfo {author} {\bibfnamefont {P.}~\bibnamefont {Malgaretti}},\
	}\bibfield  {title} {\enquote {\bibinfo {title} {Globulelike {C}onformation
				and {E}nhanced {D}iffusion of {A}ctive {P}olymers},}\ }\href {\doibase
		10.1103/PhysRevLett.121.217802} {\bibfield  {journal} {\bibinfo  {journal}
			{Phys. Rev. Lett.}\ }\textbf {\bibinfo {volume} {121}},\ \bibinfo {pages}
		{217802} (\bibinfo {year} {2018})}\BibitemShut {NoStop}%
	\bibitem [{\citenamefont {Hakim}\ and\ \citenamefont
		{Silberzan}(2017)}]{hakim}%
	\BibitemOpen
	\bibfield  {author} {\bibinfo {author} {\bibfnamefont {V.}~\bibnamefont
			{Hakim}}\ and\ \bibinfo {author} {\bibfnamefont {P.}~\bibnamefont
			{Silberzan}},\ }\bibfield  {title} {\enquote {\bibinfo {title} {Collective
				cell migration: {A} physics perspective},}\ }\href {\doibase
		10.1088/1361-6633/aa65ef} {\bibfield  {journal} {\bibinfo  {journal} {Rep.
				Prog. Phys.}\ }\textbf {\bibinfo {volume} {80}},\ \bibinfo {pages} {076601}
		(\bibinfo {year} {2017})}\BibitemShut {NoStop}%
	\bibitem [{\citenamefont {Howard}\ and\ \citenamefont {Hyman}(2007)}]{howard}%
	\BibitemOpen
	\bibfield  {author} {\bibinfo {author} {\bibfnamefont {J.}~\bibnamefont
			{Howard}}\ and\ \bibinfo {author} {\bibfnamefont {A.~A.}\ \bibnamefont
			{Hyman}},\ }\bibfield  {title} {\enquote {\bibinfo {title} {Microtubule
				polymerases and depolymerases},}\ }\href@noop {} {\bibfield  {journal}
		{\bibinfo  {journal} {Curr. Opin. Cell Biol.}\ }\textbf {\bibinfo {volume}
			{19}},\ \bibinfo {pages} {31--35} (\bibinfo {year} {2007})}\BibitemShut
	{NoStop}%
	\bibitem [{\citenamefont {Paul}\ \emph {et~al.}(2020)\citenamefont {Paul},
		\citenamefont {Majumder}, \citenamefont {Das},\ and\ \citenamefont
		{Janke}}]{paul2}%
	\BibitemOpen
	\bibfield  {author} {\bibinfo {author} {\bibfnamefont {S.}~\bibnamefont
			{Paul}}, \bibinfo {author} {\bibfnamefont {S.}~\bibnamefont {Majumder}},
		\bibinfo {author} {\bibfnamefont {S.~K.}\ \bibnamefont {Das}}, \ and\
		\bibinfo {author} {\bibfnamefont {W.}~\bibnamefont {Janke}},\ }\bibfield
	{title} {\enquote {\bibinfo {title} {Effect of alignment activity on the
				collapse kinetics of a flexible polymer},}\ }\href@noop {} {\bibfield
		{journal} {\bibinfo  {journal} {Leipzig preprint}\ } (\bibinfo {year}
		{2020})}\BibitemShut {NoStop}%
	\bibitem [{\citenamefont {Reddy}\ and\ \citenamefont
		{Thirumalai}(2017)}]{reddy}%
	\BibitemOpen
	\bibfield  {author} {\bibinfo {author} {\bibfnamefont {G.}~\bibnamefont
			{Reddy}}\ and\ \bibinfo {author} {\bibfnamefont {D.}~\bibnamefont
			{Thirumalai}},\ }\bibfield  {title} {\enquote {\bibinfo {title} {Collapse
				precedes folding in denaturant-dependent assembly of ubiquitin},}\
	}\href@noop {} {\bibfield  {journal} {\bibinfo  {journal} {J. Phys. Chem. B}\
		}\textbf {\bibinfo {volume} {121}},\ \bibinfo {pages} {995--1009} (\bibinfo
		{year} {2017})}\BibitemShut {NoStop}%
	\bibitem [{\citenamefont {Shi}\ \emph {et~al.}(2018)\citenamefont {Shi},
		\citenamefont {Liu}, \citenamefont {Hyeon},\ and\ \citenamefont
		{Thirumalai}}]{shi2018}%
	\BibitemOpen
	\bibfield  {author} {\bibinfo {author} {\bibfnamefont {G.}~\bibnamefont
			{Shi}}, \bibinfo {author} {\bibfnamefont {L.}~\bibnamefont {Liu}}, \bibinfo
		{author} {\bibfnamefont {C.}~\bibnamefont {Hyeon}}, \ and\ \bibinfo {author}
		{\bibfnamefont {D.}~\bibnamefont {Thirumalai}},\ }\bibfield  {title}
	{\enquote {\bibinfo {title} {Interphase human chromosome exhibits out of
				equilibrium glassy dynamics},}\ }\href@noop {} {\bibfield  {journal}
		{\bibinfo  {journal} {Nat. Commun.}\ }\textbf {\bibinfo {volume} {9}},\
		\bibinfo {pages} {3161} (\bibinfo {year} {2018})}\BibitemShut {NoStop}%
	\bibitem [{\citenamefont {Kaiser}\ and\ \citenamefont
		{L\"owen}(2014)}]{kaiser1}%
	\BibitemOpen
	\bibfield  {author} {\bibinfo {author} {\bibfnamefont {A.}~\bibnamefont
			{Kaiser}}\ and\ \bibinfo {author} {\bibfnamefont {H.}~\bibnamefont
			{L\"owen}},\ }\bibfield  {title} {\enquote {\bibinfo {title} {Unusual
				swelling of a polymer in a bacterial bath},}\ }\href {\doibase
		10.1063/1.4891095} {\bibfield  {journal} {\bibinfo  {journal} {J. Chem.
				Phys.}\ }\textbf {\bibinfo {volume} {141}},\ \bibinfo {pages} {044903}
		(\bibinfo {year} {2014})}\BibitemShut {NoStop}%
	\bibitem [{\citenamefont {Chaki}\ and\ \citenamefont
		{Chakrabarti}(2019)}]{chaki}%
	\BibitemOpen
	\bibfield  {author} {\bibinfo {author} {\bibfnamefont {S.}~\bibnamefont
			{Chaki}}\ and\ \bibinfo {author} {\bibfnamefont {R.}~\bibnamefont
			{Chakrabarti}},\ }\bibfield  {title} {\enquote {\bibinfo {title} {Enhanced
				diffusion, swelling, and slow reconfiguration of a single chain in
				non-gaussian active bath},}\ }\href {\doibase 10.1063/1.5086152} {\bibfield
		{journal} {\bibinfo  {journal} {J. Chem. Phys.}\ }\textbf {\bibinfo {volume}
			{150}},\ \bibinfo {pages} {094902} (\bibinfo {year} {2019})}\BibitemShut
	{NoStop}%
	\bibitem [{\citenamefont {Samanta}\ and\ \citenamefont
		{Chakrabarti}(2016)}]{samanta_16}%
	\BibitemOpen
	\bibfield  {author} {\bibinfo {author} {\bibfnamefont {N.}~\bibnamefont
			{Samanta}}\ and\ \bibinfo {author} {\bibfnamefont {R.}~\bibnamefont
			{Chakrabarti}},\ }\bibfield  {title} {\enquote {\bibinfo {title} {Chain
				reconfiguration in active noise},}\ }\href {\doibase
		10.1088/1751-8113/49/19/195601} {\bibfield  {journal} {\bibinfo  {journal}
			{J. Phys. A: Math. Theor.}\ }\textbf {\bibinfo {volume} {49}},\ \bibinfo
		{pages} {195601} (\bibinfo {year} {2016})}\BibitemShut {NoStop}%
	\bibitem [{\citenamefont {Halperin}\ and\ \citenamefont
		{Goldbart}(2000)}]{halperin}%
	\BibitemOpen
	\bibfield  {author} {\bibinfo {author} {\bibfnamefont {A.}~\bibnamefont
			{Halperin}}\ and\ \bibinfo {author} {\bibfnamefont {P.~M.}\ \bibnamefont
			{Goldbart}},\ }\bibfield  {title} {\enquote {\bibinfo {title} {Early stages
				of homopolymer collapse},}\ }\href {\doibase 10.1103/PhysRevE.61.565}
	{\bibfield  {journal} {\bibinfo  {journal} {Phys. Rev. E}\ }\textbf {\bibinfo
			{volume} {61}},\ \bibinfo {pages} {565--573} (\bibinfo {year}
		{2000})}\BibitemShut {NoStop}%
	\bibitem [{\citenamefont {Majumder}\ and\ \citenamefont
		{Janke}(2015)}]{majumder1}%
	\BibitemOpen
	\bibfield  {author} {\bibinfo {author} {\bibfnamefont {S.}~\bibnamefont
			{Majumder}}\ and\ \bibinfo {author} {\bibfnamefont {W.}~\bibnamefont
			{Janke}},\ }\bibfield  {title} {\enquote {\bibinfo {title} {Cluster
				coarsening during polymer collapse: Finite-size scaling analysis},}\ }\href
	{\doibase 10.1209/0295-5075/110/58001} {\bibfield  {journal} {\bibinfo
			{journal} {Europhys. Lett.}\ }\textbf {\bibinfo {volume} {110}},\ \bibinfo
		{pages} {58001} (\bibinfo {year} {2015})}\BibitemShut {NoStop}%
	\bibitem [{\citenamefont {Majumder}\ \emph {et~al.}(2017)\citenamefont
		{Majumder}, \citenamefont {Zierenberg},\ and\ \citenamefont
		{Janke}}]{majumder3}%
	\BibitemOpen
	\bibfield  {author} {\bibinfo {author} {\bibfnamefont {S.}~\bibnamefont
			{Majumder}}, \bibinfo {author} {\bibfnamefont {J.}~\bibnamefont
			{Zierenberg}}, \ and\ \bibinfo {author} {\bibfnamefont {W.}~\bibnamefont
			{Janke}},\ }\bibfield  {title} {\enquote {\bibinfo {title} {Kinetics of
				polymer collapse: {E}ffect of temperature on cluster growth and aging},}\
	}\href {\doibase 10.1039/C6SM02197B} {\bibfield  {journal} {\bibinfo
			{journal} {Soft Matter}\ }\textbf {\bibinfo {volume} {13}},\ \bibinfo {pages}
		{1276--1290} (\bibinfo {year} {2017})}\BibitemShut {NoStop}%
	\bibitem [{\citenamefont {Majumder}\ \emph {et~al.}(2020)\citenamefont
		{Majumder}, \citenamefont {Christiansen},\ and\ \citenamefont
		{Janke}}]{majumder4}%
	\BibitemOpen
	\bibfield  {author} {\bibinfo {author} {\bibfnamefont {S.}~\bibnamefont
			{Majumder}}, \bibinfo {author} {\bibfnamefont {H.}~\bibnamefont
			{Christiansen}}, \ and\ \bibinfo {author} {\bibfnamefont {W.}~\bibnamefont
			{Janke}},\ }\bibfield  {title} {\enquote {\bibinfo {title} {Understanding
				nonequilibrium scaling laws governing collapse of a polymer},}\ }\href@noop
	{} {\bibfield  {journal} {\bibinfo  {journal} {Eur. Phys. J. B}\ }\textbf
		{\bibinfo {volume} {93}},\ \bibinfo {pages} {1--19} (\bibinfo {year}
		{2020})}\BibitemShut {NoStop}%
	\bibitem [{\citenamefont {Christiansen}\ \emph {et~al.}(2017)\citenamefont
		{Christiansen}, \citenamefont {Majumder},\ and\ \citenamefont
		{Janke}}]{christiansen}%
	\BibitemOpen
	\bibfield  {author} {\bibinfo {author} {\bibfnamefont {H.}~\bibnamefont
			{Christiansen}}, \bibinfo {author} {\bibfnamefont {S.}~\bibnamefont
			{Majumder}}, \ and\ \bibinfo {author} {\bibfnamefont {W.}~\bibnamefont
			{Janke}},\ }\bibfield  {title} {\enquote {\bibinfo {title} {Coarsening and
				aging of lattice polymers: Influence of bond fluctuations},}\ }\href
	{\doibase 10.1063/1.4991667} {\bibfield  {journal} {\bibinfo  {journal} {J.
				Chem. Phys.}\ }\textbf {\bibinfo {volume} {147}},\ \bibinfo {pages} {094902}
		(\bibinfo {year} {2017})}\BibitemShut {NoStop}%
	\bibitem [{\citenamefont {Byrne}\ \emph {et~al.}(1995)\citenamefont {Byrne},
		\citenamefont {Kiernan}, \citenamefont {Green},\ and\ \citenamefont
		{Dawson}}]{byrne}%
	\BibitemOpen
	\bibfield  {author} {\bibinfo {author} {\bibfnamefont {A.}~\bibnamefont
			{Byrne}}, \bibinfo {author} {\bibfnamefont {P.}~\bibnamefont {Kiernan}},
		\bibinfo {author} {\bibfnamefont {D.}~\bibnamefont {Green}}, \ and\ \bibinfo
		{author} {\bibfnamefont {K.~A.}\ \bibnamefont {Dawson}},\ }\bibfield  {title}
	{\enquote {\bibinfo {title} {Kinetics of homopolymer collapse},}\ }\href
	{\doibase 10.1063/1.469437} {\bibfield  {journal} {\bibinfo  {journal} {J.
				Chem. Phys.}\ }\textbf {\bibinfo {volume} {102}},\ \bibinfo {pages}
		{573--577} (\bibinfo {year} {1995})}\BibitemShut {NoStop}%
	\bibitem [{\citenamefont {Bunin}\ and\ \citenamefont {Kardar}(2015)}]{bunin}%
	\BibitemOpen
	\bibfield  {author} {\bibinfo {author} {\bibfnamefont {G.}~\bibnamefont
			{Bunin}}\ and\ \bibinfo {author} {\bibfnamefont {M.}~\bibnamefont {Kardar}},\
	}\bibfield  {title} {\enquote {\bibinfo {title} {Coalescence {M}odel for
				{C}rumpled {G}lobules {F}ormed in {P}olymer {C}ollapse},}\ }\href {\doibase
		10.1103/PhysRevLett.115.088303} {\bibfield  {journal} {\bibinfo  {journal}
			{Phys. Rev. Lett.}\ }\textbf {\bibinfo {volume} {115}},\ \bibinfo {pages}
		{088303} (\bibinfo {year} {2015})}\BibitemShut {NoStop}%
	\bibitem [{\citenamefont {Guo}\ \emph {et~al.}(2011)\citenamefont {Guo},
		\citenamefont {Liang},\ and\ \citenamefont {Wang}}]{guo}%
	\BibitemOpen
	\bibfield  {author} {\bibinfo {author} {\bibfnamefont {J.}~\bibnamefont
			{Guo}}, \bibinfo {author} {\bibfnamefont {H.}~\bibnamefont {Liang}}, \ and\
		\bibinfo {author} {\bibfnamefont {Z.-G.}\ \bibnamefont {Wang}},\ }\bibfield
	{title} {\enquote {\bibinfo {title} {Coil-to-globule transition by
				dissipative particle dynamics simulation},}\ }\href {\doibase
		10.1063/1.3604812} {\bibfield  {journal} {\bibinfo  {journal} {J. Chem.
				Phys.}\ }\textbf {\bibinfo {volume} {134}},\ \bibinfo {pages} {244904}
		(\bibinfo {year} {2011})}\BibitemShut {NoStop}%
	\bibitem [{\citenamefont {Milchev}\ \emph {et~al.}(2001)\citenamefont
		{Milchev}, \citenamefont {Bhattacharya},\ and\ \citenamefont
		{Binder}}]{milchev}%
	\BibitemOpen
	\bibfield  {author} {\bibinfo {author} {\bibfnamefont {A.}~\bibnamefont
			{Milchev}}, \bibinfo {author} {\bibfnamefont {A.}~\bibnamefont
			{Bhattacharya}}, \ and\ \bibinfo {author} {\bibfnamefont {K.}~\bibnamefont
			{Binder}},\ }\bibfield  {title} {\enquote {\bibinfo {title} {Formation of
				block copolymer micelles in solution:  {A} {M}onte {C}arlo study of chain
				length dependence},}\ }\href {\doibase 10.1021/ma000645j} {\bibfield
		{journal} {\bibinfo  {journal} {Macromolecules}\ }\textbf {\bibinfo {volume}
			{34}},\ \bibinfo {pages} {1881--1893} (\bibinfo {year} {2001})}\BibitemShut
	{NoStop}%
	\bibitem [{\citenamefont {Weeks}\ \emph {et~al.}(1971)\citenamefont {Weeks},
		\citenamefont {Chandler},\ and\ \citenamefont {Andersen}}]{wca_71}%
	\BibitemOpen
	\bibfield  {author} {\bibinfo {author} {\bibfnamefont {J.~D.}\ \bibnamefont
			{Weeks}}, \bibinfo {author} {\bibfnamefont {D.}~\bibnamefont {Chandler}}, \
		and\ \bibinfo {author} {\bibfnamefont {H.~C.}\ \bibnamefont {Andersen}},\
	}\bibfield  {title} {\enquote {\bibinfo {title} {Role of repulsive forces in
				determining the equilibrium structure of simple liquids},}\ }\href {\doibase
		10.1063/1.1674820} {\bibfield  {journal} {\bibinfo  {journal} {J. Chem.
				Phys.}\ }\textbf {\bibinfo {volume} {54}},\ \bibinfo {pages} {5237--5247}
		(\bibinfo {year} {1971})}\BibitemShut {NoStop}%
	\bibitem [{\citenamefont {Frenkel}\ and\ \citenamefont {Smit}(2002)}]{frenkel}%
	\BibitemOpen
	\bibfield  {author} {\bibinfo {author} {\bibfnamefont {D.}~\bibnamefont
			{Frenkel}}\ and\ \bibinfo {author} {\bibfnamefont {B.}~\bibnamefont {Smit}},\
	}\href@noop {} {\emph {\bibinfo {title} {Understanding Molecular Simulation:
				From Algorithms to Applications}}}\ (\bibinfo  {publisher} {Academic Press,
		San Diego},\ \bibinfo {year} {2002})\BibitemShut {NoStop}%
	\bibitem [{\citenamefont {Schnabel}\ \emph {et~al.}(2009)\citenamefont
		{Schnabel}, \citenamefont {Bachmann},\ and\ \citenamefont
		{Janke}}]{schnabel}%
	\BibitemOpen
	\bibfield  {author} {\bibinfo {author} {\bibfnamefont {S.}~\bibnamefont
			{Schnabel}}, \bibinfo {author} {\bibfnamefont {M.}~\bibnamefont {Bachmann}},
		\ and\ \bibinfo {author} {\bibfnamefont {W.}~\bibnamefont {Janke}},\
	}\bibfield  {title} {\enquote {\bibinfo {title} {Elastic {L}ennard-{J}ones
				polymers meet clusters: Differences and similarities},}\ }\href {\doibase
		10.1063/1.3223720} {\bibfield  {journal} {\bibinfo  {journal} {J. Chem.
				Phys.}\ }\textbf {\bibinfo {volume} {131}},\ \bibinfo {pages} {124904}
		(\bibinfo {year} {2009})}\BibitemShut {NoStop}%
	\bibitem [{\citenamefont {Milchev}\ and\ \citenamefont
		{Binder}(1994)}]{milchev_94}%
	\BibitemOpen
	\bibfield  {author} {\bibinfo {author} {\bibfnamefont {A.}~\bibnamefont
			{Milchev}}\ and\ \bibinfo {author} {\bibfnamefont {K.}~\bibnamefont
			{Binder}},\ }\bibfield  {title} {\enquote {\bibinfo {title} {Anomalous
				diffusion and relaxation of collapsed polymer chains},}\ }\href {\doibase
		10.1209/0295-5075/26/9/006} {\bibfield  {journal} {\bibinfo  {journal}
			{Europhys. Lett.}\ }\textbf {\bibinfo {volume} {26}},\ \bibinfo {pages}
		{671--676} (\bibinfo {year} {1994})}\BibitemShut {NoStop}%
	\bibitem [{\citenamefont {Paul}\ and\ \citenamefont {Das}(2014)}]{paul_epl}%
	\BibitemOpen
	\bibfield  {author} {\bibinfo {author} {\bibfnamefont {S.}~\bibnamefont
			{Paul}}\ and\ \bibinfo {author} {\bibfnamefont {S.~K.}\ \bibnamefont {Das}},\
	}\bibfield  {title} {\enquote {\bibinfo {title} {Dynamics of clustering in
				freely cooling granular fluid},}\ }\href {\doibase
		10.1209/0295-5075/108/66001} {\bibfield  {journal} {\bibinfo  {journal}
			{Europhys. Lett.}\ }\textbf {\bibinfo {volume} {108}},\ \bibinfo {pages}
		{66001} (\bibinfo {year} {2014})}\BibitemShut {NoStop}%
\end{thebibliography}
\end{document}